\documentclass[reprint,superscriptaddress,showpacs,showkeys,amsmath,amssymb,aps,nofootinbib,preprintnumbers]{revtex4-1}

\usepackage{graphicx}
\usepackage[mathlines]{lineno}
\usepackage[english]{babel}
\usepackage{color}
\usepackage{ulem}

\def \nutau {$\nu_\tau~$}
\def \vavrg {$\langle V \rangle ~$}

\def \pao {Pierre Auger Observatory}

\def\lapproxeq{\lower .7ex\hbox{$\;\stackrel{\textstyle
<}{\sim}\;$}}
\def\gapproxeq{\lower .7ex\hbox{$\;\stackrel{\textstyle
>}{\sim}\;$}}

\begin{document}

\preprint{Published in PRD as doi:10.1103/PhysRevD.91.092008}

\title{An improved limit to the diffuse flux of ultra-high energy neutrinos \\ 
from the Pierre Auger Observatory}

\author{A.~Aab}
\affiliation{Universit\"{a}t Siegen, Siegen, 
Germany}
\author{P.~Abreu}
\affiliation{Laborat\'{o}rio de Instrumenta\c{c}\~{a}o e F\'{\i}sica 
Experimental de Part\'{\i}culas - LIP and  Instituto Superior 
T\'{e}cnico - IST, Universidade de Lisboa - UL, 
Portugal}
\author{M.~Aglietta}
\affiliation{Osservatorio Astrofisico di Torino  (INAF), 
Universit\`{a} di Torino and Sezione INFN, Torino, 
Italy}
\author{E.J.~Ahn}
\affiliation{Fermilab, Batavia, IL, 
USA}
\author{I.~Al Samarai}
\affiliation{Institut de Physique Nucl\'{e}aire d'Orsay (IPNO), 
Universit\'{e} Paris 11, CNRS-IN2P3, Orsay, 
France}
\author{I.F.M.~Albuquerque}
\affiliation{Universidade de S\~{a}o Paulo, Instituto de F\'{\i}sica, 
S\~{a}o Paulo, SP, 
Brazil}
\author{I.~Allekotte}
\affiliation{Centro At\'{o}mico Bariloche and Instituto Balseiro 
(CNEA-UNCuyo-CONICET), San Carlos de Bariloche, 
Argentina}
\author{P.~Allison}
\affiliation{Ohio State University, Columbus, OH, 
USA}
\author{A.~Almela}
\affiliation{Universidad Tecnol\'{o}gica Nacional - Facultad 
Regional Buenos Aires, Buenos Aires, 
Argentina}
\affiliation{Instituto de Tecnolog\'{\i}as en Detecci\'{o}n y 
Astropart\'{\i}culas (CNEA, CONICET, UNSAM), Buenos Aires, 
Argentina}
\author{J.~Alvarez Castillo}
\affiliation{Universidad Nacional Aut\'{o}noma de M\'{e}xico, M\'{e}xico,
 D.F., 
M\'{e}xico}
% ...
\collaboration{The Pierre Auger Collaboration}
\email{{\tt auger\_spokespersons@fnal.gov}}
\noaffiliation

\begin{abstract}
Neutrinos in the cosmic ray flux with energies near 1 EeV and above are detectable 
with the Surface Detector array of the Pierre Auger Observatory. We report here on searches
through Auger data from 1 January 2004 until 20 June 2013. No neutrino candidates were found, 
yielding a limit to the diffuse flux of ultra-high energy neutrinos that challenges the 
Waxman-Bahcall bound predictions.  
Neutrino identification is attempted using the broad time-structure 
of the signals expected in the SD stations, and is efficiently done for neutrinos of all flavors 
interacting in the atmosphere at large zenith angles, as well as for ``Earth-skimming" neutrino 
interactions in the case of tau neutrinos. In this paper the searches 
for downward-going neutrinos in the zenith angle bins 
$60^\circ-75^\circ$ and $75^\circ-90^\circ$ as well as for upward-going neutrinos, 
are combined to give a single limit. 
The $90\%$ C.L. single-flavor limit to the diffuse flux 
of ultra-high energy neutrinos with an $E^{-2}$ spectrum in the energy range 
$1.0 \times 10^{17}$ eV - $2.5 \times 10^{19}$ eV 
is $E_\nu^2 dN_\nu/dE_\nu < 6.4 \times 10^{-9}~ {\rm GeV~ cm^{-2}~ s^{-1}~ sr^{-1}}$.
\end{abstract}

\pacs{95.55.Vj, 95.85.Ry, 98.70.Sa} 

\keywords{Ultra-high-energy cosmic rays and neutrinos, high-energy showers, 
ground detector arrays, Pierre Auger Observatory}

\maketitle

\section{Introduction}
\def\linenumberfont{\normalfont\tiny\itshape\sffamily}

The flux of ultra-high energy cosmic rays (UHECRs) above $\sim 5\times 10^{19}$ eV 
is known to be suppressed with respect to that extrapolated from lower energies. 
This feature has been seen in the UHECR spectrum \cite{HiRes_spectrum,Auger_spectrum},
with the position of the break being compatible with the Greisen-Zatsepin-Kuzmin (GZK) effect \cite{GZK},  
i.e. the interaction of UHECRs with the cosmic microwave background (CMB) radiation.
However, other explanations are possible, most prominently a scenario 
where the limiting energy of the UHECR sources is being observed \cite{Allard_2012}. 
Key to distinguishing between these two scenarios is the determination
of the composition of the UHECRs \cite{Auger_Xmax,TA_Xmax}, with the second scenario predicting
increasing fractions of primaries heavier than protons as energy increases \cite{Allard_2012}. 

Above $\sim 5\times 10^{19}$ eV cosmic-ray protons interact with CMB photons
and produce ultra-high energy {\it cosmogenic} neutrinos of energies 
typically $1/20$ of the proton energy \cite{BZ}. Their fluxes are uncertain 
and at EeV energies they depend mostly on the evolution with redshift $z$ of
the unknown sources of UHECRs, and on their spectral features at injection. 
Protons typically produce more neutrinos than heavier primaries do \cite{Ave_GZK,Kotera_GZK},
so measurement of the neutrino flux gives information on the nature of the primaries.
In this respect the observation of UHE neutrinos can provide further hints 
on the dominant scenario of UHECR production \cite{Kotera_GZK}, as well as on the evolution with $z$
of their sources which can help in their identification \cite{Stanev,Kotera_GZK}. 
 
UHE neutrinos are also expected to be produced 
in the decay of charged pions created in the interactions of cosmic rays with 
matter and/or radiation at their potential sources, 
such as Gamma-Ray Bursts or Active Galactic Nuclei 
among others \cite{Becker_PhysRep}. 
In fact, at tens of EeV, neutrinos may be the only direct probe 
of the sources of UHECRs at distances farther 
than $\sim 100$ Mpc.

A breakthrough in the field was the recent detection with the 
IceCube experiment of three neutrinos of energies just above 1 PeV, 
including a 2 PeV event which is the highest-energy neutrino 
interaction ever observed, followed by tens of others above $\sim 30$ TeV 
representing a $\sim5.7~\sigma$ excess
above atmospheric neutrino background \cite{IceCube_PRL14}. 
The measured flux is close to the Waxman-Bahcall upper bound to the UHE neutrino flux \cite{WB},
although with a steeper spectrum \cite{IceCube_PRD15}.

In the EeV energy range, i.e. about three orders of magnitude above the most energetic 
neutrinos detected in IceCube, neutrinos 
have so far escaped detection by existing experiments. 
These can be detected with a variety of techniques \cite{Veronique},
among them with arrays of particle detectors at ground.

In this work we report on the search for EeV neutrinos in data taken 
with the Surface Detector array (SD) of the Pierre Auger Observatory \cite{Auger_SD}. 
A blind scan of data from 1 January 2004 up to 20 June 2013 has yielded no neutrino 
candidates and an updated and stringent limit to the diffuse flux of UHE neutrino flux
has been obtained. 

\section{Searching for UHE neutrinos in Auger}

The concept for identification of neutrinos is rather simple.
While protons, heavier nuclei, and even photons interact shortly
after entering the atmosphere, neutrinos can initiate showers quite
deep in the atmosphere. 
At large zenith angles the atmosphere is thick enough 
so that the electromagnetic component of nucleonic cosmic rays 
gets absorbed and the shower front at ground level is dominated by muons  
(``old'' shower front). On the other hand, showers induced
by neutrinos deep in the atmosphere have a considerable amount of electromagnetic 
component at the ground (``young'' shower front).
The Surface Detector array (SD) of the Pierre Auger Observatory is not directly sensitive 
to the muonic and electromagnetic components of the shower separately, nor 
to the depth at which the shower is initiated.  
In the ${\sim} 1600$ 
water-Cherenkov stations of the SD of the Pierre Auger Observatory, spread over
an area of ${\sim} 3000~{\rm km^2}$, separated by $1.5~{\rm km}$ 
and arranged in a triangular grid, the signals produced by the 
passage of shower particles are digitised with 
Flash Analog to Digital Converters (FADC) with 25 ns resolution. 
This allows us to distinguish narrow signals 
in time induced by inclined showers initiated high in the atmosphere, 
from the broad signals expected in inclined showers initiated close to the ground.

Applying this simple idea, with the SD of the Pierre Auger Observatory~\cite{Auger_SD} 
we can efficiently detect inclined showers and search for two types of neutrino-induced 
showers at energies above about 1 EeV:

\begin{enumerate}

\item 
Earth-skimming (ES) showers induced by tau neutrinos ($\nu_\tau$)
that travel in a slightly upward direction with respect to ground.
\nutau can skim the Earth's crust and interact relatively close to the surface 
inducing a tau lepton which escapes the Earth and decays in flight in the atmosphere,
close to the SD. 

Typically, only Earth-skimming \nutau-induced showers with zenith 
angles $90^\circ < \theta < 95^\circ$ may be identified. 

\item 
Showers initiated by any neutrino flavor moving down at large angles with respect
to the vertical that interact 
in the atmosphere close 
to the surface detector array
through charged-current (CC) or neutral-current
(NC) interactions.  We include here showers induced
by \nutau interacting in the mountains surrounding the \pao. Although  
this latter process is exactly equivalent to the
``Earth-skimming'' mechanism, it is included in this class because such showers 
are also going downwards. In the following we will refer to all these types
of showers as ``downward-going" (DG) $\nu$-induced showers.

With the aid of Monte Carlo simulations we have established that this search can be performed 
efficiently as long as it is restricted to showers with zenith 
angles $\theta > 60^\circ$. Due to the characteristics
of these showers depending on the zenith angle, the search in this 
channel was performed in two angular subranges: 
(a) ``low" zenith angle (DGL) corresponding to $60^\circ < \theta < 75^\circ$ 
and (b) ``high" zenith angle (DGH) with $75^\circ < \theta < 90^\circ$. 

\end{enumerate}

\subsection{General procedure} 

The identification of potential neutrino-induced showers 
is based on first selecting those events that arrive in 
rather inclined directions,
and then selecting among them those with FADC traces that are spread in time,
indicative of the early stage of development of the shower and 
a clear signature of a deeply interacting neutrino triggering the SD.

First of all, events occurring during periods of data acquisition instabilities
\cite{Auger_trigger} are excluded.
For the remaining events the FADC traces of the triggered stations are first
``cleaned'' to remove accidental signals \cite{ES} induced mainly by random atmospheric muons
arriving closely before or after the shower front.
These muons are typically produced in lower energy showers (below
the energy threshold of the SD of the Auger Observatory) that arrive by chance
in coincidence with the triggering shower. 
A procedure to select the stations participating in the event described
in \cite{ES,DGH} is then applied, with the event accepted if 
the number of accepted stations $N_{\rm st}$ is at least 
three (four) in the Earth-skimming (downward-going) selections.

From the pattern (footprint) of stations at ground
a length $L$ along the arrival direction
of the event and a width $W$ perpendicular to it
characterizing the shape of the footprint are extracted \cite{ES}. 
The ratio $L/W\sim 1$ in vertical events, increasing
gradually as the zenith angle increases.
Very inclined events typically have elongated patterns on the
ground along the direction of arrival and hence  
large values of $L/W$. 
A cut in $L/W$ is therefore a good discriminator of inclined events.
Another indication of inclined events is given by the apparent speed
$V$ of the trigger from a station $i$ to a station $j$, averaged over
all pairs $(i,j)$ of stations in the event. This observable denoted as \vavrg is obtained 
from the distance
between the stations after projection along $L$ 
and from the difference in trigger times of the stations. 
In vertical showers \vavrg exceeds the speed of light since 
all triggers occur at roughly the same time, while in very inclined events 
\vavrg is concentrated around the speed of light. 
Moreover its Root-Mean-Square (RMS($V$)) value is small.
For downward-going events only, a cut on the reconstructed zenith angle 
$\theta_{\rm rec}$ is applied \cite{DGH}. 

Once inclined showers are selected the next step is to identify young showers. 
A Time-over-Threshold (ToT) trigger\footnote{This trigger 
is intended to select sequences of small signals in the FADC traces spread in time.
It requires at least 13 bins in 120 FADC bins
of a sliding window of 3 $\mu$s above a
threshold of $0.2~I_{VEM}^{peak}$ (the peak value of the signal expected for a vertical
muon crossing the station), in coincidence in 2 out of 3 PMTs \cite{Auger_trigger}.} 
is usually present in SD stations with signals extended in time, 
while narrow signals induce other local triggers. 
Also the Area-over-Peak ratio (AoP), defined as
the ratio of the integral of the FADC trace to its peak value, 
normalized to the average signal produced by a single muon,
provides an estimate of the spread-in-time of the traces, 
and serves as an observable to discriminate broad
from narrow shower fronts. 
In particular, a cut on AoP allows the rejection of background signals induced by
inclined hadronic showers, in which the muons and their electromagnetic products
are concentrated within a short time interval, exhibiting AoP values close to the one
measured in signals induced by isolated muons. These observables are 
used by themselves in the search for $\nu$ candidates, or combined in a linear 
Fisher-discriminant polynomial depending on the
selection as described later in this work. 

As a general procedure and to optimize the numerical values of the cuts and tune 
the algorithms needed to separate neutrino-induced showers
from the much larger background of hadronic showers, we divided
the whole data sample (1 January 2004 - 20 June 2013) 
into two parts (excluding periods of array instability). 
A selection dependent fraction of the data $\sim 20\%$, 
along with Monte Carlo simulations of UHE neutrinos, 
is dedicated to define the selection algorithm,
the most efficient observables and the value of the cuts on them.
These data are assumed to be overwhelmingly constituted of background showers.
The applied procedure is conservative because the
presence of neutrinos in the training data would result in a more severe 
definition of the selection criteria.
The remaining fraction of data is not used until the selection procedure
is established, and then it is ``unblinded" to search for neutrino candidates. 
We used real data to train the selections instead of Monte Carlo simulations
of hadronic showers, 
the primary reason being that the detector simulation may not account 
for all possible detector fluctuations that may induce 
events that constitute a background to UHE neutrinos, while they are
contained in data.
It is important to remark that this is the same selection procedure
and training period as in previous publications \cite{ES,DGH},
which is applied in this work to a larger data set.

Regarding the Monte Carlo simulations, the phase space of the neutrino 
showers reduces to three variables:
the neutrino energy $E_\nu$, the incidence zenith angle $\theta$ 
and the interaction depth $D$ in the atmosphere for downward-going neutrinos,
or the altitude $h_c$ of the $\tau$ decay above ground
in the case of Earth-skimming neutrinos.
Showers were simulated
with energies from $\log(E_{\tau}/{\rm eV}) = 17$ to $20.5$ in steps of $0.5$,
zenith angles from $90.1^{\circ}$ to $95.9^{\circ}$ in steps of 0.01 rad (ES) 
and from $60^{\circ}$ to $90^{\circ}$ in steps of 0.05 rad (DG).
The values of $h_c$ range from 0 to 2500 m (in steps of 100 m)
whereas $D$ is uniformly distributed along the shower axis
in steps of 100 g cm$^{-2}$.

We have described the general procedure to search for Earth-skimming \nutau and 
downward-going $\nu$-induced showers. However the two searches (ES and DG) 
differ in several aspects
that we describe in the following sections.

\subsection{Earth-skimming (ES) neutrinos} 

With Monte Carlo simulations of UHE \nutau 
propagating inside the Earth, we have established
that $\tau$ leptons above the energy threshold
of the SD are efficiently produced only at 
zenith angles between $90^\circ$ and $95^\circ$. 
For this reason, in the Earth-skimming analysis 
we place very restrictive cuts 
to select only quasi-horizontal showers 
with largely elongated footprints:
$L/W>5$ and  \vavrg $\in [0.29, 0.31]~{\rm m~ns^{-1}}$
with RMS($V$)$<0.08~{\rm m~ns^{-1}}$ 
(see Table~\ref{tab:cuts})\footnote{
The axis of Earth-skimming showers travelling in
the upward direction does not intersect the ground,
contrary to the case for downward-going showers. 
For this reason, we exploit the properties of the
footprint generated by the shower particles that
deviate laterally from the shower axis and trigger
the SD water-Cherenkov stations.}.

\begin{figure}[!t]
\centering
\includegraphics[width=8.5cm]{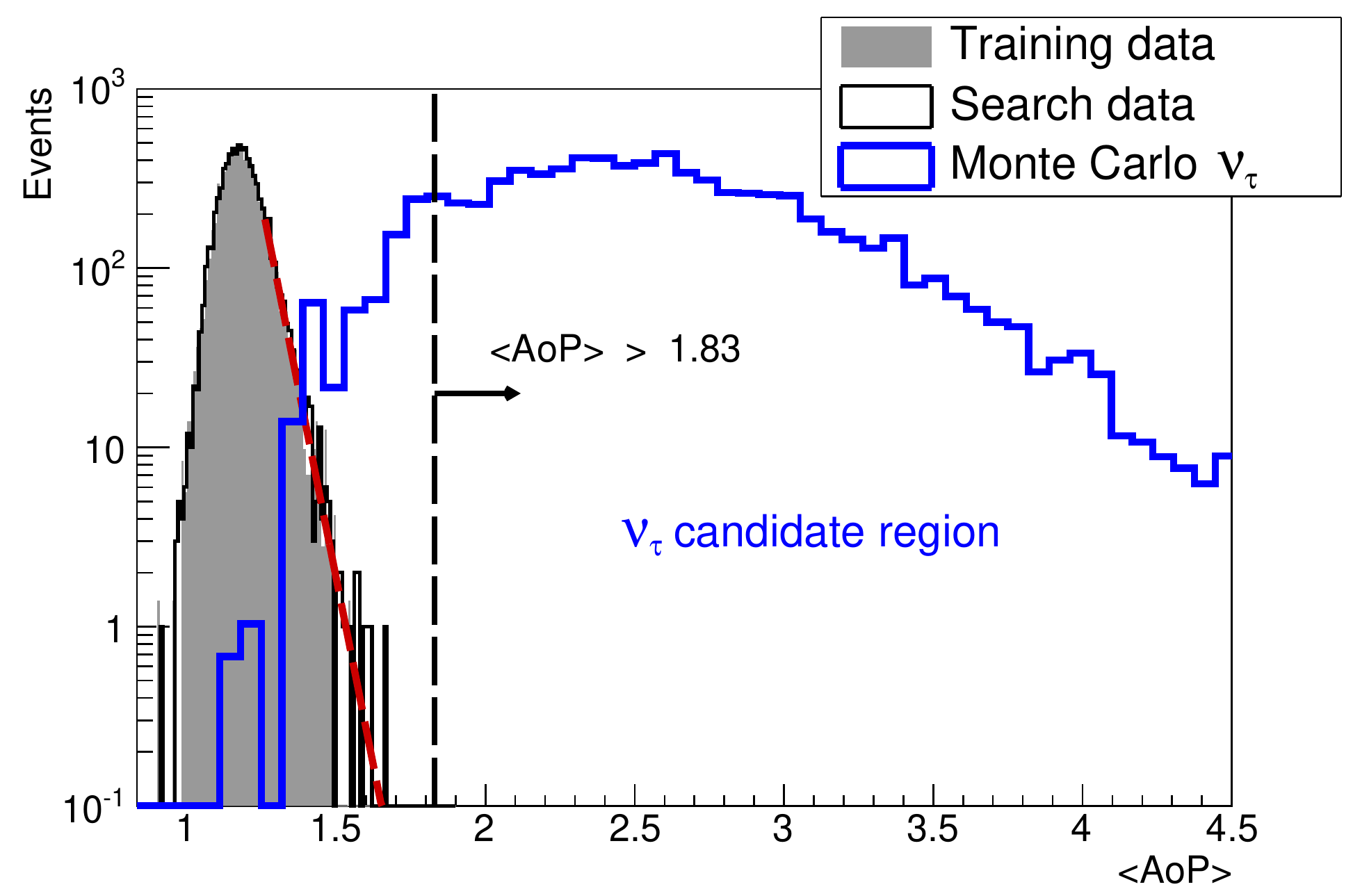}
\vskip -3mm
\caption{
Distributions of $\langle$AoP$\rangle$ (the variable used to identifiy neutrinos
in the ES selection for data after 1 June 2010)
after applying the inclined shower selection in Table~\ref{tab:cuts}.
Gray-filled histogram: the data in the training period. 
Black histogram: data 
in the search period. These two distributions are normalised to the same number 
of events for comparison purposes. Blue histogram: 
simulated ES $\nu_\tau$ events. The dashed vertical line represents 
the cut on $\langle$AoP$\rangle$ $>$ 1.83 above which a data event is regarded as a neutrino candidate.
An exponential fit to the tail of the distribution of training data is also shown 
as a red dashed line (see text for explanation).
}
\label{fig:AoP}
\end{figure}

In the ES selection, the neutrino identification variables include the fraction 
of stations with ToT trigger and having AoP$>1.4$ for data prior to 31 May 2010 \cite{ES}. 
This fraction is required to be above $60\%$ of the triggered stations in the event.
The final choice of the values of these cuts was 
made by requiring zero background events in the training data sample,
corresponding to $1\%$ of the events recorded up to that date.
For data beyond 1 June 2010 a new methodology and 
a new set of efficient selection criteria was established
based on an improved and enlarged library of ES 
simulated $\nu_\tau$ events 
and on a larger period of training data. 
In particular, we used the average value of AoP ($\langle$AoP$\rangle$) 
over all the triggered stations in the event as the main observable to 
discriminate between hadronic showers and ES neutrinos. 
The new methodology allows us to place the 
value of the cut on $\langle$AoP$\rangle$ 
using the tail of its distribution 
as obtained in real data
(which was seen to be consistent with an exponential shape as shown in Fig.~\ref{fig:AoP}). 
This tail was fitted and extrapolated to find the value of the cut corresponding 
to less than 1 expected event per 50 yr on the full SD array. 
As a result, an event is tagged as a neutrino candidate if $\langle$AoP$\rangle> 1.83$
(see Table~\ref{tab:cuts} and Fig.~\ref{fig:AoP}). 
The new methodology is not applied to the data prior to 31 May 2010
since that data period was already unblinded to search for UHE neutrinos 
under the older cuts \cite{ES}. 

Roughly $\sim 95\%$ of the simulated 
inclined $\nu_\tau$ events producing $\tau$ leptons above the energy threshold of the SD 
are kept after the cut on $\langle$AoP$\rangle$. The search for neutrinos is clearly not 
limited by background in this channel.  

\subsection{Downward-going (DG) neutrinos} 

In the high zenith angle range of the downward-going analysis (DGH)
the values of the cuts to select inclined events are obtained in 
Monte Carlo simulations of events with $\theta>75^\circ$. Due 
to the larger angular range compared to Earth-skimming $\nu_\tau$,
less stringent criteria are applied, namely $L/W>3$,
\vavrg$<0.313~{\rm m~ns^{-1}}$, RMS($V$)/\vavrg$<0.08$
plus a further requirement that the reconstructed zenith 
angle $\theta_{\rm rec}>75^\circ$ (see \cite{DGH} and Table \ref{tab:cuts}
for full details).

In the low zenith angle range (DGL) corresponding to $60^\circ < \theta < 75^\circ$,
$L/W$, \vavrg and RMS($V$)/\vavrg are less efficient in selecting 
inclined events than the reconstructed zenith angle $\theta_{\rm rec}$,
and for this reason only a cut on $\theta_{\rm rec}$ is applied, namely
$58.5^\circ < \theta_{\rm rec} < 76.5^\circ$,
which includes some allowance to account for the resolution in the angular reconstruction 
of the simulated neutrino events. 

After the inclined shower selection is peformed, the discrimination power is optimized 
with the aid of the multi-variate Fisher discriminant method \cite{Fisher}. 
A linear combination of observables is constructed which optimizes the separation between
background hadronic inclined showers occuring during 
the downward-going training period, and Monte Carlo
simulated $\nu$-induced showers. The method requires as input 
a set of observables. For that purpose 
we use variables depending on the dimensionless Area-over-Peak (AoP) 
observable -- as defined above -- of the FADC traces. 

\begin{figure}[t!]
\centering
\includegraphics[width=8.5cm]{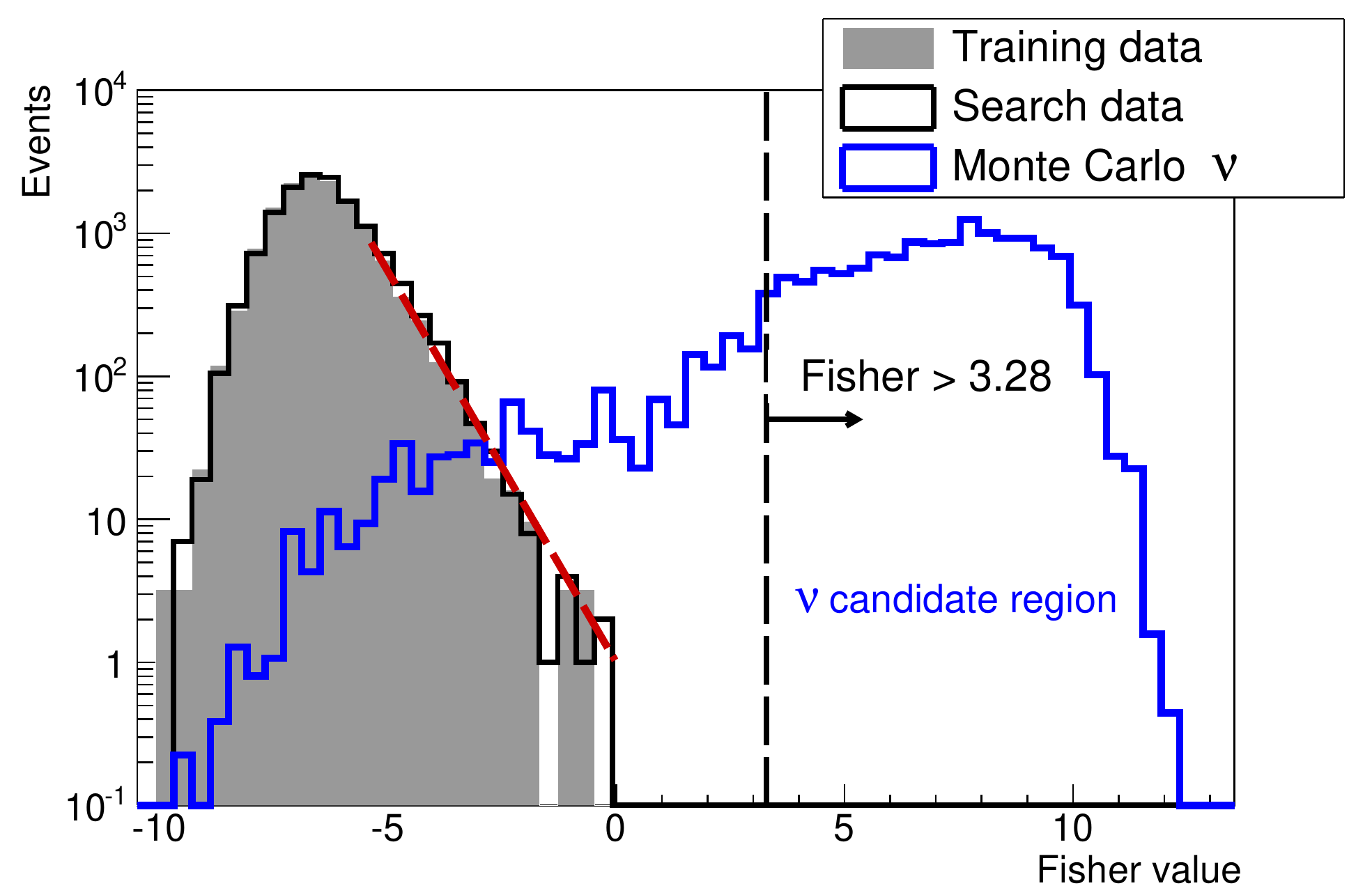}
\vskip -3mm
\caption{
Distributions of the Fisher variable $\cal F$ 
in inclined events selected by the ``Inclined Showers" DGH criteria 
in Table~\ref{tab:cuts}, before applying the ``Young Showers" cuts. 
In particular the distribution of events with number of triggered
tanks $7\leq~N_{\rm st}~\leq11$ is shown. 
Gray-filled histogram: data 
in the training period corresponding to $\sim23\%$ of the whole
data sample between 1 January 2004 and 20 June 2013. 
Black line: data 
in the search period. The distributions are normalised to the same number 
of events for comparison purposes. Blue line: 
simulated DGH $\nu$ events. The dashed vertical line represents 
the cut on ${\cal F} > 3.28$ above which a data event is regarded as a neutrino candidate.
The red dashed line represents an exponential fit to the tail of the training distribution (see text
for explanation). 
}
\label{fig:Fisher}
\end{figure}

In the DGH channel, due to the inclination of the shower the electromagnetic component 
is less attenuated at the locations of the stations that are first hit by a deep inclined shower
({\it early stations}) than in the stations that are hit last ({\it late stations}). From Monte Carlo 
simulations of $\nu-$induced showers with $\theta > 75^\circ$ we have established that in the first 
few early stations the typical AoP values range between 3 and 5, while AoP tends to be closer to 1 in 
the late stations. Based on this simple observation and as already reported in \cite{DGH},
we have found a good discrimination when the following ten variables are used to construct
the linear Fisher discriminant variable $\cal F$: the
AoP and (AoP)$^2$ of the four stations that trigger first in each event,
the product of the four AoPs, and a global parameter that
measures the asymmetry between the average AoP of the early
stations and those triggering last in the event 
(see \cite{DGH} for further details and Table~\ref{tab:cuts}).
 
The selection of neutrino candidates in the zenith angle range 
$60^\circ < \theta < 75^\circ$ (DGL) is more challenging since 
the electromagnetic component of background hadronic showers 
at ground increases as the zenith angle decreases because 
the shower crosses less atmosphere before reaching the detector level. 
Out of all triggered stations of an event in this angular 
range, the ones closest to the shower core exhibit the highest
discrimination power in terms of AoP. 
In fact it has been observed in Monte Carlo simulations that 
the first triggered stations can still contain some electromagnetic component 
for background events and, for this reason, it is not desirable to use them
for discrimination purposes. 
The last ones, even if they are triggered only by muons from a background
hadronic shower, can exhibit large values of AoP because they are far from 
the core where muons are known to arrive with a larger spread in time.
Based on the information from Monte Carlo simulations, 
the variables used in the Fisher discriminant analysis
are the individual AoP of the four or five stations (depending on the 
zenith angle) closest to the core, and their product \cite{DGL}.
In the DGL analysis it is also required that at least $75\%$ of the triggered stations 
closest to the core have a ToT local trigger \cite{DGL}.

Once the Fisher discriminant $\cal F$ is defined, the next
step is to define a numerical value ${\cal F}_{\rm cut}$ 
that efficiently separates neutrino candidates
from regular hadronic showers. 
As was done for the variable $\langle$AoP$\rangle$
in the Earth-skimming analysis, ${\cal F}_{\rm cut}$
was fixed using the tail of the distribution of $\cal F$ in real data, 
which is consistent with an exponential shape in all cases. An example
is shown in Fig.~\ref{fig:Fisher}. The tail 
was fitted and extrapolated to find the value of ${\cal F}_{\rm cut}$ 
corresponding to less than 1 expected event per 50 yr on the full SD array \cite{DGH,DGL}.
Roughly $\sim 85\%$ ($\sim 60\%$) of the simulated 
inclined $\nu$ events are kept after the cut on the Fisher variable 
in the DGH (DGL) selections. 
The smaller efficiencies for the identification of 
neutrinos in the DGL selection are due to the more stringent criteria in the angular bin $\theta \in (60^\circ, 75^\circ)$ 
needed to reject the larger contamination from cosmic-ray induced showers.

\begin{table*}[!t]
\begin{center}
\renewcommand{\arraystretch}{1.4}
\begin{tabular}{l|c|c|c}
\hline
~Selection & Earth-skimming (ES)           & Downward-going                        & Downward-going                       \\
          &                               & {\it high} angle (DGH)                & {\it low} angle (DGL)                \\
\hline 
~Flavours $\&$ Interactions~~ & $\nu_\tau$ CC & $\nu_e,~\nu_\mu~,\nu_\tau$ CC $\&$ NC &  $\nu_e,~\nu_\mu~,\nu_\tau$ CC $\&$ NC \\

~Angular range & $\theta>90^\circ$             & $\theta \in (75^\circ, 90^\circ)$ & $\theta \in (60^\circ, 75^\circ)$ \\

~N$^\circ$ of Stations ($N_{\rm st}$) & $N_{\rm st}$ $\geq$ 3   & $N_{\rm st}$ $\geq$ 4 & $N_{\rm st}$ $\geq$ 4 \\
\hline 
          & $-$                             & $\theta_{\rm rec}>$ 75$^{\circ}$   &   $\theta_{\rm rec}\in (58.5^\circ,~76.5^{\circ})$\\
~Inclined  & $L/W > 5$                                         & $L/W > 3$ & $-$ \\
~Showers   & $\langle V\rangle\in (0.29,~0.31)~{\rm m~ns^{-1}}$ & $\langle V \rangle~<~0.313~{\rm m~ns^{-1}}$ & $-$ \\
          & RMS($V$)$~<~0.08~{\rm m~ns^{-1}}$                 & RMS($V$)/$\langle V\rangle<0.08$ & $-$ \\
\hline 
          & ~Data: 1 January 2004 - 31 May 2010~ &                                     & $\geq 75\%$ of stations close to   \\
          & $\geq 60\%$ of stations with  &                                      & ~shower core with ToT trigger~       \\
~Young     & ToT trigger \& AoP $>$ 1.4    & ~Fisher discriminant based~         &                          \&        \\
~Showers   & Data: 1 June 2010 - 20 June 2013   & on AoP of {\it early} stations  & Fisher discriminant based  \\
          & $\langle {\rm AoP} \rangle> 1.83$  &                                 & on AoP of {\it early} stations  \\
          & AoP$_{\rm min}$ $>$ 1.4 if $N_{\rm st}$=3   &                                 & close to shower core        \\
\hline
\end{tabular}
\end{center}
\vskip -3mm
\caption{
Observables and numerical values of cuts applied to select {\it inclined} 
and {\it young} showers for Earth-skimming and downward-going neutrinos. See text for explanation.} 
\label{tab:cuts}
\end{table*}

\section{Data unblinding and exposure calculation}

\subsection{Data unblinding}

No events survived when the Earth-skimming and downward-going selection criteria
explained above and summarized in Table~\ref{tab:cuts} are applied 
blindly to the data collected between 1 January 2004 and 20 June 2013.
For each selection the corresponding training periods are excluded
from the search. 
After the unblinding we tested the compatibility of the distributions
of discriminating observables in the search and training samples. 
Examples are shown in Fig.~\ref{fig:AoP} for the $\langle{\rm AoP}\rangle$
variable in the Earth-skimming analysis, and in Fig.~\ref{fig:Fisher} for the 
Fisher variable in the DGH analysis. In particular
fitting the tails of the corresponding distributions to an exponential,
we obtained compatible parameters within 1$~\sigma$ statistical uncertainties. 

\subsection{Exposure calculation}

\subsubsection{Neutrino identification efficiencies}

The selection criteria in Table~\ref{tab:cuts}, were also applied 
to neutrino-induced showers simulated with Monte Carlo, and 
the identification efficiencies
$\epsilon_{\rm ES},~\epsilon_{\rm DGH},~\epsilon_{\rm DGL}$ for 
each channel - defined as the fraction of simulated events 
passing the cuts - were obtained.

A large set of Monte Carlo simulations of neutrino-induced
showers was performed for this purpose, covering the whole
parameter space where the efficiency is expected to be sizeable.
In the case of Earth-skimming \nutau induced showers, 
the efficiency depends on the energy of the 
emerging $\tau$ leptons $E_\tau$, on the zenith angle $\theta$ and on the altitude of the
decay point of the $\tau$ above ground. These efficiencies are averaged over azimuthal 
angle and the $\tau$ decay channels. 
The maximum efficiency that can be reached
is $82.6\%$, the $17.4\%$ remaining corresponds to the channel 
in which the $\tau$ decays into a $\mu$ which is unlikely to 
produce a detectable shower close to ground. 
In the case of downward-going neutrinos the identification efficiency
depends on neutrino flavor, type of interaction (CC or NC), neutrino 
energy $E_\nu$, zenith angle $\theta$, and distance $D$ measured
from ground along the shower axis at which the neutrino is forced to interact
in the simulations.

The identification efficiencies depend also on time, through the changing configuration 
of the SD array that was growing steadily since 2004 up to 2008, and because  
the fraction of working stations - although typically 
above $95\%$ - is changing continuously with time.
Also the continuous monitoring of the array reveals
a slight evolution with time of the optical properties of 
the water-Cherenkov stations (see below).  
Although the number of working stations and their status
are monitored every second and as a consequence the SD configuration
is known with very good accuracy at any instant of time,
in practice, to avoid having to cope with an impractically large number of configurations,
different strategies were devised to calculate in an accurate and less time-consuming
manner the actual identification efficiencies (as explained in \cite{ES,DGH,DGL}). 

The evolution of the optical properties of the water-Cherenkov stations 
was taken into account in an effective way in the calculation of 
the exposure. The main effect of this evolution is a decrease with 
time of the decay-time of the light as obtained from the monitoring
data that revealed a continuous decrease of  
$\sim 10\%$ from 2004 until 
the end of the data period used in this work (20 June 2013).  
This induces a reduction of the AoP and, as a consequence,
the trigger efficiency changes with time.
These changes were accounted for in the  
calculation of the exposure by dividing the whole data set 
into three separate periods and assuming that in each of them the decay-time 
of the light in the tank remained approximately constant as seen in data. 
A conservative approach was adopted by choosing constant values 
of the light decay-time below the actual curve in the three periods. 

\subsubsection{Combination of selections}

In previous publications \cite{ES,DGH,DGL} the fraction of $\nu$-induced Monte Carlo events 
identified as neutrino candidates was obtained by applying each particular set of 
selection criteria (ES, DGH, DGL) only to its corresponding set of simulated showers 
(ES, DGH or DGL). In this work the fraction of selected events is further increased by applying 
the three sets of criteria to each sample of simulated showers (ES, DGH, DGL) regardless of channel. 
With this procedure the fraction of identified Monte Carlo events is enhanced as, for instance, 
an ES simulated shower induced by a \nutau might not fulfill the requirements of the ES selection, 
but might still pass the DGH or DGL criteria, and hence contribute to the fraction of identified
events. The enhancement in the fraction of events when applying this ``combined" analysis
depends on the particular set of Monte Carlo simulations. For instance applying the 
three criteria to the DGH Monte Carlo sample identifies a fraction of neutrino events 
$\sim 1.25$ larger than when the DGH criteria are applied alone, the enhancement
coming mainly from events with 3 stations rejected by the DGH criteria but accepted by ES.
The application of the three criteria to the ES Monte Carlo
sample however results in a smaller enhancement $\sim 1.04$. 

\subsubsection{Exposure calculation}
 
For downward-going neutrinos, once the efficiencies $\epsilon_{\rm DG}(E_\nu,\theta,D,t)$ 
are obtained, the calculation 
of the exposure involves folding them with the SD array aperture and the $\nu$ interaction probability
at a depth $D$ for a neutrino energy $E_\nu$.  
This calculation also includes the possibility that downward-going $\nu_\tau$ interact
with the mountains surrounding the Observatory. 
Integrating over the parameter space except for $E_\nu$ and in time over the search periods 
and summing over all the interaction channels yields the exposure \cite{DGH,DGL}. 

In the Earth-skimming channel, $\epsilon_{\rm ES}(E_\tau,\theta,X_d)$
are also folded with the aperture, with the probability density function of a tau emerging from the
Earth with energy $E_\tau$ (given a neutrino with energy $E_\nu$ crossing an amount of Earth
determined by the zenith angle $\theta$), as well as with the probability that 
the $\tau$ decays at an altitude $h_c$ \cite{ES}. An integration over the whole parameter
space except for $E_\nu$ and time gives the exposure \cite{ES}. 

\begin{figure}[!t]
\centering
\includegraphics[width=9.0cm]{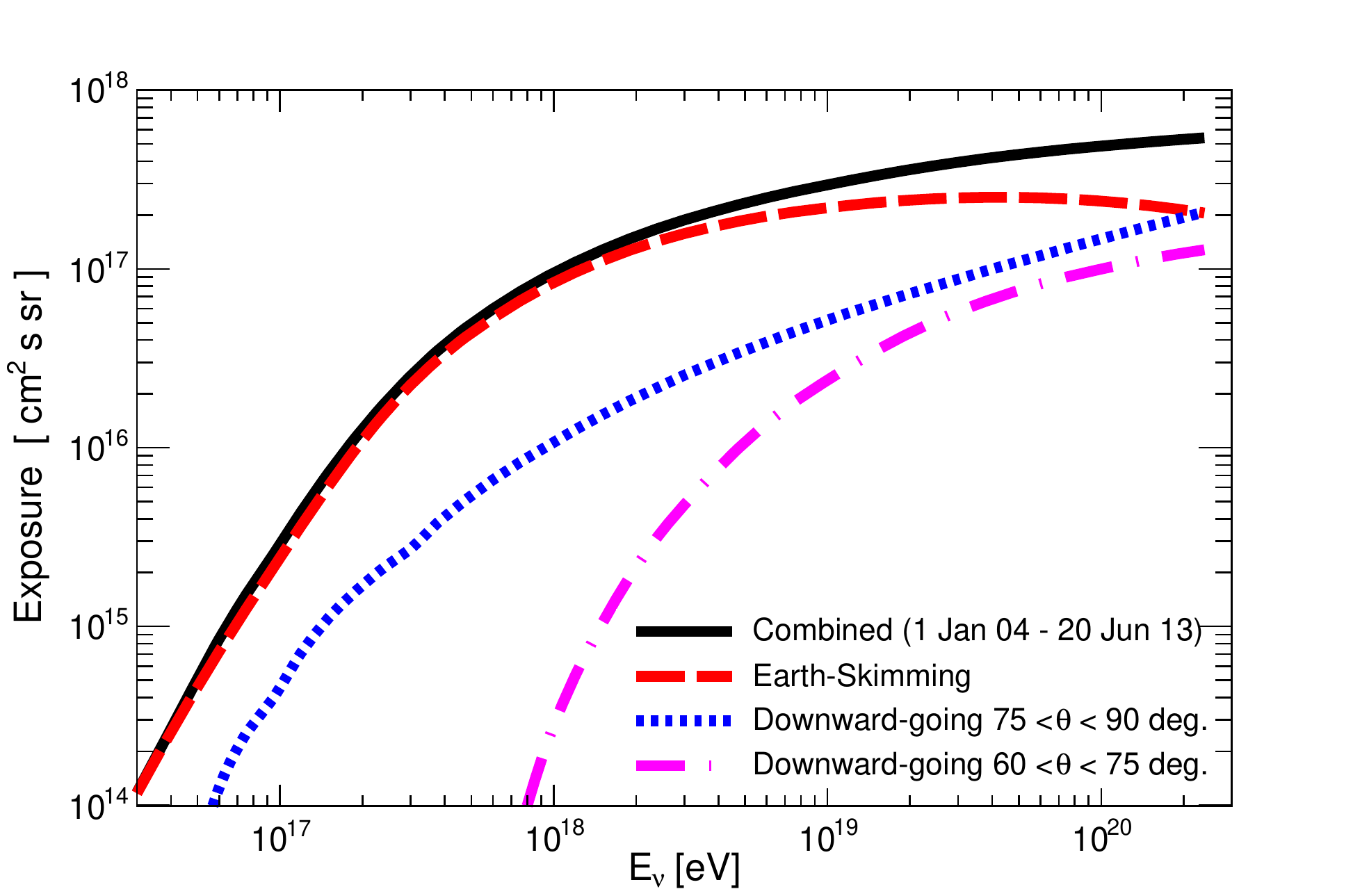}
\vskip -3mm
\caption{
Combined exposure of the SD of the Pierre Auger Observatory
(1 January 2004 - 20 June 2013)
as a function of neutrino energy after applying the three sets 
of selection criteria in Table~\ref{tab:cuts}
to Monte Carlo simulations of UHE neutrinos (see text for explanation). Also shown are 
the individual exposures corresponding to each of the three selections.
For the downward-going channels the exposure represents the sum over the
three neutrino flavors as well as CC and NC interactions.
For the Earth-Skimming channel, only $\nu_\tau$ CC interactions are relevant.
}
\label{fig:exposure}
\end{figure}

The exposures ${\cal E}_{\rm ES}$, ${\cal E}_{\rm DGH}$ 
and ${\cal E}_{\rm DGL}$ obtained for the search periods of each selection 
are plotted in Fig.~\ref{fig:exposure} along with their sum $\cal E_{\rm tot}$. 
The exposure to Earth-skimming neutrinos is higher than that to 
downward-going neutrinos,  
partially due to the longer search period in the Earth-skimming 
analysis, and partially due to the much larger neutrino conversion
probability in the denser target of the Earth's crust compared to the 
atmosphere. The larger number of neutrino flavors and interaction channels 
that can be identified in the DGH and DGL analysis, 
as well as the broader angular range $60^\circ<\theta<90^\circ$ 
partly compensates the dominance of the ES channel.  
The ES exposure flattens and then falls above $\sim 10^{19}$ eV as there is an
increasing probability that the $\tau$ decays high in the atmosphere producing a 
shower not triggering the array, or even that the $\tau$ escapes the atmosphere before decaying.
At the highest energies the DGH exposure dominates.
The DGL exposure is the smallest of the three, mainly due to the more stringent
criteria needed to apply to get rid of the larger background nucleonic showers
in the zenith angle bin $60^\circ<\theta<75^\circ$.

The relative contributions of the three channels 
to the total expected event rate 
for a differential flux behaving with energy as $dN_\nu(E_\nu)/dE_\nu \propto E_\nu^{-2}$ 
are ES:DGH:DGL$\sim$0.84:0.14:0.02 respectively, where the event rate is obtained as:
\begin{equation} 
N_{\rm evt}={\int_{E_{\nu}}~\frac{dN_\nu}{dE_\nu}(E_\nu)~{\cal E}_{\rm tot}(E_\nu)~dE_\nu}
\label{eq:Nevt}
\end{equation}
%

\subsection{Systematic uncertainties}

Several sources of systematic uncertainty have been considered.
Some of them are directly related to the Monte Carlo simulation of the showers,
i.e., generator of the neutrino interaction either in the Earth or in the 
atmosphere, parton distribution function, air shower development, and
hadronic model. 

Other uncertainties have to do with the limitations on the
theoretical models needed to obtain the interaction
cross-section or the $\tau$ energy loss at high energies.
In the Earth-skimming analysis the model of energy
loss for the $\tau$ is the dominant source of uncertainty, since it determines the 
energy of the emerging $\tau$s after propagation in the Earth; the impact of this on the downward-going analysis is 
much smaller since $\tau$ energy losses are only relevant for \nutau interacting
in the mountains, a channel that is estimated to contribute only $\sim15\%$ to the DGH exposure \cite{DGH}. 

The uncertainty on the shower simulation, that stems mainly from the different shower propagation codes
and hadronic interaction models that can be used to model the high energy collisions in the shower, 
contributes significantly in the ES and DG channels. 

The presence of mountains around the Observatory -- which would increase the target for 
neutrino interactions in both cases -- is explicitly simulated and 
accounted for when obtaining the exposure of the SD to downward-going neutrino-induced showers,
and as a consequence does not contribute directly to the systematic uncertainties.  
However, it is not accounted for in the Earth-skimming channel 
and instead we take the topography around
the Observatory as a source of systematic uncertainty.

In the three channels the procedure to incorporate the systematic uncertainties
is the same. Different combinations of the various
sources of systematic uncertainty render different values of the exposure
and a {\it systematic uncertainty band} of relative deviation from a 
reference exposure (see below) can be constructed for each channel and for each source
of systematic uncertainty.
For a given source of uncertainty the edges of the ES, DGH and DGL bands are 
weighted by the relative importance of each channel as given before and
added linearly or quadratically depending on the source of uncertainty.  
In Table~\ref{tab:sys} we give the dominant sources of systematic uncertainty
and their corresponding combined uncertainty bands obtained in this way.
The combined uncertainty band is then incorporated in the value of the limit itself 
through a semi-Bayesian extension \cite{Conrad} of the Feldman-Cousins 
approach \cite{Feldman-Cousins}.

In the calculation of the reference exposure 
the $\nu$-nucleon interaction in the atmosphere for DG neutrinos (including
CC and NC channels) is simulated with HERWIG \cite{HERWIG}.
In the case of $\nu_\tau$ CC interactions, a dedicated, fast and flexible
code is used to simulate the $\tau$ lepton propagation in the Earth and/or
in the atmosphere. The $\tau$ decay is performed with the TAUOLA package \cite{TAUOLA}.
In all cases we adopted the $\nu$-nucleon cross-section in \cite{Cooper-Sarkar}.
In a second step, the AIRES code \cite{AIRES} is used to simulate the propagation in the atmosphere of the
particles produced in the high energy $\nu$ interaction or in the $\tau$ lepton decay.
The types, energies, momenta and times of the particles reaching the SD level are obtained.
The last stage is the simulation of the SD response (PMT signals and FADC traces).
This involves a modification of the ``standard" sampling procedure in 
\cite{Billoir_unthinning} to regenerate 
particles in the SD stations from the ``thinned" air shower simulation output, 
that was tailored to the
highly inclined showers involved in the search for neutrinos.   
Light production and propagation inside the station is based on GEANT4 \cite{GEANT4}
with the modifications to account for the evolution of the light decay-time explained above.
These two latter changes roughly compensate each other, with the net result
being a few percent decrease of the exposure with respect to that obtained
with the standard thinning procedure and a constant average value of 
the light decay-time.

\begin{table}[!t]
\begin{center}
\renewcommand{\arraystretch}{1.3}
\begin{tabular}{ l c }
\hline
 Source of  systematic                    & ~~~~Combined uncertainty band    \\
\hline
 Simulations                              & ~~~~$\sim$~+4\%, -3\%               \\

 $\nu$ cross section \& $\tau$ E-loss     & ~~~~$\sim$~+34\%, -28\%                                 \\

 Topography      	                  & ~~~~$\sim$~+15\%, 0\%                 \\

\hline

 Total                                    & ~~~~$\sim$~+37\%, -28\%             \\

\hline
\end{tabular}
\end{center}
\vskip -3mm
\caption{
Main sources of systematic uncertainties and their
corresponding combined uncertainty bands (see text for details)
representing the effect on the event rate defined in Eq.~(\ref{eq:Nevt}).  
The uncertainty due to ``Simulations" includes: interaction generator, shower simulation, hadronic model, 
thinning and detector simulator.
The uncertainty due to ``$\tau$ energy-loss" affects the ES channel and 
also the DGH but only to $\nu_\tau$ with $\theta\gtrsim 88^\circ$ going through 
the mountains surrounding the Pierre Auger Observatory. However it does not affect the DGL channel.
The topography around the Observatory is not accounted for in the ES channel 
and is taken as a systematic uncertainty that would increase the event rate. 
}
\label{tab:sys}
\end{table}

\section{Results}

\begin{figure}[!t]
\centering
\includegraphics[width=9.0cm]{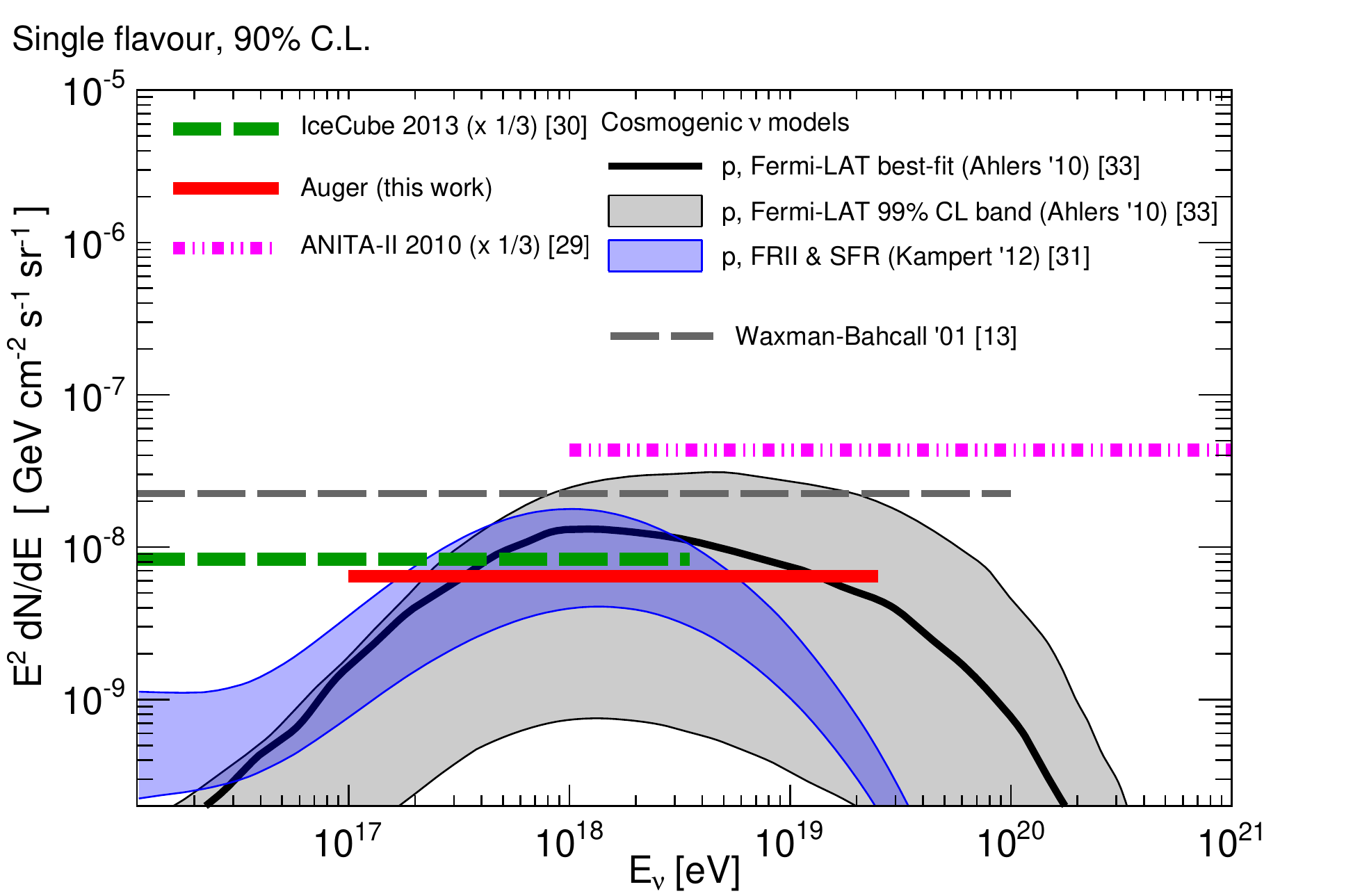}
\includegraphics[width=9.0cm]{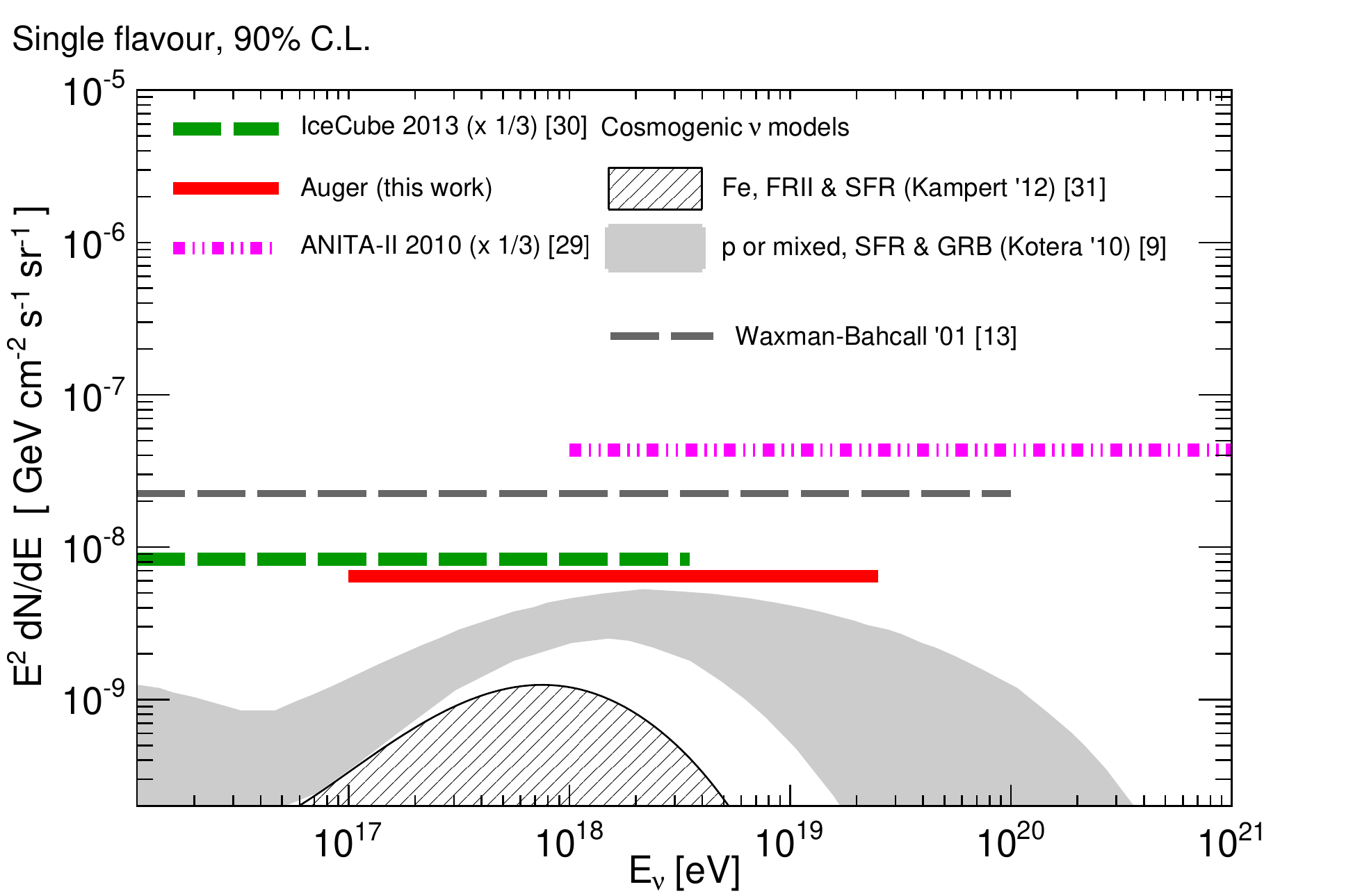}
\vskip -3mm
\caption{ 
Top panel: Upper limit (at 90$\%$ C.L.)
to the normalization of the diffuse flux of UHE neutrinos 
as given in Eqs.~(\ref{eq:limit}) and (\ref{eq:k90}),
from the Pierre Auger Observatory. 
We also show the corresponding limits from ANITAII \cite{ANITAII} 
and IceCube \cite{IceCube_latest_limit} experiments,
along with expected fluxes for several cosmogenic neutrino models that assume pure protons as primaries \cite{Ahlers_GZK,Kampert_GZK}
as well as the Waxman-Bahcall bound \cite{WB}.
All limits and fluxes converted to single flavor.
We used $N_{\rm up}=2.39$ in Eq.~(\ref{eq:limit}) 
to obtain the limit (see text for details).
Bottom panel: Same as top panel, but showing several cosmogenic 
neutrino models that assume heavier nuclei as primaries, 
either pure iron \cite{Kampert_GZK} or mixed primary compositions \cite{Kotera_GZK}.
}
\label{fig:limits}
\end{figure}

\begin{figure}[!t]
\centering
\includegraphics[width=9.0cm]{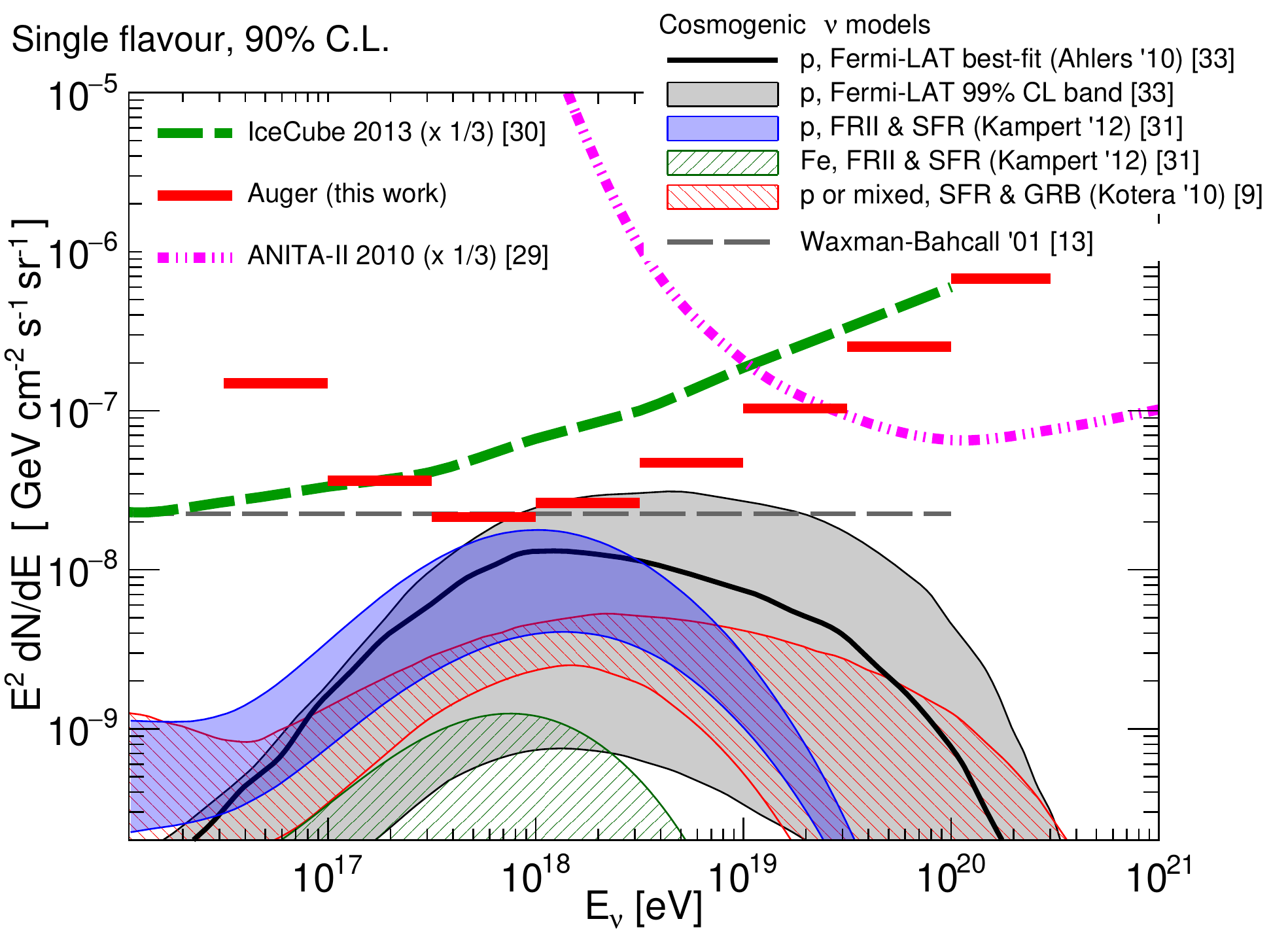}
\vskip -3mm
\caption{ 
Upper limit to the normalization of the diffuse flux
of UHE neutrinos (at 90$\%$ C.L. and in bins of width 0.5 
in $\log_{10}E_\nu$ - see text for details) from the Pierre Auger Observatory
(straight steps). We also show the corresponding limits from ANITAII \cite{ANITAII}
(dot-dashed line) and IceCube \cite{IceCube_latest_limit} (dashed line) experiments
(with appropriate normalizations to take into account the energy bin
width, and to convert to single flavor), along with expected  fluxes
for several cosmogenic neutrino models \cite{Ahlers_GZK,Kampert_GZK,Kotera_GZK}
as well as the Waxman-Bahcall bound \cite{WB} (all converted to single flavor).
}
\label{fig:limits2}
\end{figure}

Using the combined exposure in Fig.~\ref{fig:exposure} 
and assuming a differential neutrino flux $dN(E_\nu)/dE_\nu = k\cdot E_\nu^{-2}$ 
as well as a $\nu_e:\nu_\mu:\nu_\tau=1:1:1$ flavor ratio, an upper limit on the value of $k$ can be obtained as:
\begin{equation}
k = \frac{N_{\rm up}}{\int_{E_{\nu}}~E_{\nu}^{-2}~{\cal E}_{\rm tot}(E_\nu)~dE_\nu}.
\label{eq:limit}
\end{equation}
The actual value of the upper limit on the signal events ($N_{\rm up}$)
depends on the number of observed events ($0$ in our case) 
and expected background events (conservatively assumed to be $0$), as well
as on the confidence level required ($90\%$ C.L. in the following).
Using a semi-Bayesian extension~\cite{Conrad}
of the Feldman-Cousins approach \cite{Feldman-Cousins} 
to include the uncertainties in the exposure we obtain\footnote{To calculate 
$N_{\rm up}$ we use POLE++ \cite{Conrad}. 
The signal efficiency uncertainty is $\sim 0.19$ with an asymmetric band 
(see Table~\ref{tab:sys}). This yields a value 
of $N_{\rm up}=2.39$ slightly smaller than the nominal 2.44 of 
the Feldman-Cousins approach.} $N_{\rm up}=2.39$.  
The single-flavor $90\%$ C.L. limit is:
\begin{equation}
k_{90}~< ~6.4 \times 10^{-9}~{\rm GeV~cm^{-2}~s^{-1}~sr^{-1}}.
\label{eq:k90}
\end{equation}
The limit applies  
in the energy interval $\sim 1.0\times10^{17}~{\rm eV} - 2.5\times10^{19}~{\rm eV}$
where the cumulative number of events as a function of neutrino energy increases 
from $5\%$ to $95\%$ of the total number, i.e. where $\sim 90\%$ of the total 
event rate is expected.
It is important to remark that this is the most stringent limit obtained 
so far with Auger data, and it represents a single limit combining 
the three channels where we have searched for UHE neutrinos.
The limit to the flux normalization in Eq.~(\ref{eq:k90}) is obtained integrating the denominator 
of Eq.~(\ref{eq:limit}) in the whole energy range where Auger is sensitive to UHE neutrinos. 
This is shown in Fig.~\ref{fig:limits} , along with the $90\%$ C.L. limits from other experiments 
as well as several models of neutrino flux production (see caption for references). 
The denominator of Eq.~(\ref{eq:limit}) can also be integrated in bins of energy, 
and a limit on $k$ can also be obtained in each energy bin \cite{Anchordoqui}. 
This is displayed in Fig.~\ref{fig:limits2}  
where the energy bins have a width of $0.5$ in $\log_{10}E_\nu$, 
and where we also show the whole energy range where there is sensitivity 
to neutrinos. The limit as displayed in Fig.~\ref{fig:limits2} allows us to show 
at which energies the sensitivity of the SD of the Pierre Auger Observatory peaks.

The search period corresponds to an equivalent of $6.4$ years of a complete 
Auger SD array working continuously. The inclusion of the data from 1 June 2010 until 20 June 2013 
in the search represents an increase of a factor $\sim 1.8$ in total time quantified in terms
of equivalent full Auger years with respect to previous 
searches \cite{ES,DGH}. Further improvements in the limit come from the combination of the three
analysis into a single one, using the 
procedure explained before that enhances the fraction of identified neutrinos especially in the DGH channel. 

In Table~\ref{tab:rates} we give the expected total event rates for several models
of neutrino flux production. 

\begin{table*}[!t]
\begin{center}
\renewcommand{\arraystretch}{1.3}
\begin{tabular}{l c c} 
\hline
Diffuse flux       &  Expected number of events  & ~~~Probability of  \\
Neutrino Model     &  (1 January 2004 - 20 June 2013)     & ~~~observing $0$   \\
\hline
Cosmogenic - proton, FRII \cite{Kampert_GZK}     &  $\sim$ 4.0  & ~~~$\sim 1.8\times 10^{-2}$ \\

Cosmogenic - proton, SFR \cite{Kampert_GZK}      &  $\sim$ 0.9  & ~~~$\sim 0.4$               \\

Cosmogenic - proton, Fermi-LAT, $E_{\rm min}=10^{19}$ eV \cite{Ahlers_GZK} &  ~~~$\sim$ 3.2  & $\sim 4\times 10^{-2}$   \\

Cosmogenic - proton, Fermi-LAT, $E_{\rm min}=10^{17.5}$ eV \cite{Ahlers_GZK} &  ~~~$\sim$ 1.6  & $\sim 0.2$   \\

Cosmogenic - proton or mixed, SFR \& GRB \cite{Kotera_GZK}              &  $\sim$ 0.5 $-$ 1.4 & ~~~$\sim 0.6~-~0.2$ \\

Cosmogenic - iron, FRII \cite{Kampert_GZK}       &  $\sim$ 0.3  & ~~~$\sim 0.7$ \\

\hline

Astrophysical $\nu$ (AGN) \cite{Becker_AGN}     &  $\sim$ 7.2  & ~~~$\sim 7\times 10^{-4}$ \\

\hline

Exotic \cite{Sigl}                               &  $\sim$ 31.5  & ~~~$\sim 2\times 10^{-14}$ \\

\hline
\end{tabular}
\end{center}
\vskip -3mm
\caption{
Number of expected events $N_{\rm evt}$ in Eq.~(\ref{eq:Nevt}) for several theoretical models of UHE neutrino 
production (see Figs.~\ref{fig:limits} and \ref{fig:limits2}), given the combined exposure of the surface detector array
of the Pierre Auger Observatory plotted in Fig.~\ref{fig:exposure}.
The last column gives the Poisson probability $\exp({-N_{\rm evt}})$ of observing 0 events 
when the number of expected events is $N_{\rm evt}$ given in the second column.
}
\label{tab:rates}

\end{table*}

Several important conclusions and remarks can be stated
after inspecting Figs.~\ref{fig:limits} and \ref{fig:limits2} and Table~\ref{tab:rates}:

\begin{enumerate}

\item
The maximum sensitivity of the SD of the Auger Observatory is achieved
at neutrino energies around EeV, where most cosmogenic models
of $\nu$ production also peak (in a $E_\nu^2 \times dN/dE_\nu$ plot). 

\item
The current Auger limit is a factor $\sim4$ below the Waxman-Bahcall bound on neutrino 
production in optically thin sources \cite{WB}. The SD of the Auger Observatory
is the first air shower array to reach that level of sensitivity. 

\item
Some models of neutrino production in astrophysical sources 
such as Active Galactic Nuclei (AGN) are excluded at more than 90\% C.L.   
For the model $\#2$ shown in Fig.~14 of \cite{Becker_AGN} 
we expect ${\sim} 7$ neutrino events while none was observed.  

\item
Cosmogenic $\nu$ models that assume a pure primary proton composition injected at the sources
and strong (FRII-type) evolution of the sources are strongly disfavored by Auger data. 
An example is the upper line of the shaded band in 
Fig.~17 in \cite{Kampert_GZK} (also depicted in Figs.~\ref{fig:limits} and \ref{fig:limits2}), 
for which ${\sim} 4$ events are expected
and as consequence that flux is excluded at ${\sim}98\%$ C.L.
Models that assume a pure primary proton composition 
and use the GeV $\gamma$-ray flux observations by the 
Fermi-LAT satellite detector as an additional constraint, 
are also disfavored.
For instance for the model shown as a solid line in the bottom right panel of Fig.~5 in \cite{Ahlers_GZK}
(also depicted in Figs.~\ref{fig:limits} and \ref{fig:limits2} in this work),  
corresponding to the best-fit to the cosmic-ray spectrum as measured by HiRes,
we expect ${\sim} 3.2$ events. As a consequence that model is excluded at more than $90\%$ C.L.
For this particular model we also show in Figs.~\ref{fig:limits} and \ref{fig:limits2} the $99\%$ C.L. 
band resulting from the fitting to the HiRes spectrum down to $E_{\rm min}=10^{19}$ eV. 
The Auger limit is also approaching the solid line in the upper left panel of Fig.~5 in \cite{Ahlers_GZK},
a model that assumes extragalactic protons above $E_{\rm min}=10^{17.5}$ eV \cite{Berezinsky_dip_model}, 
for which $\sim 1.6$ events are expected (see Table~\ref{tab:rates}).
The Auger direct limits on cosmogenic neutrinos 
are also constraining part of the region indirectly bounded by Fermi-LAT observations.

\item
The current Auger limit is less restrictive with the cosmogenic neutrino 
models represented by the gray shaded area in the bottom panel of Fig.~\ref{fig:limits}
(${\sim} 0.5$ to ${\sim} 1.4$ events are expected as shown in Table~\ref{tab:rates})
which brackets the lower fluxes predicted under a 
range of assumptions for the composition of the
primary flux (protons or mixed), source evolution and model for
the transition from Galactic to extragalactic cosmic-rays \cite{Kotera_GZK}
The same remark applies to models
that assume pure-iron composition at the sources. 
A 10-fold increase in the current exposure will be
needed to reach the most optimistic predictions
of cosmogenic neutrino fluxes if the primaries are 
pure iron, clearly out of the range of the current 
configuration of the Auger Observatory. 

\item 
A large range of exotic models of neutrino production \cite{Sigl} are excluded 
with C.L. larger than $99\%$.

\item
In IceCube, neutrino fluxes in the 30 TeV to 2 PeV energy range 
have shown a ${\sim} 5.7\sigma$ excess compared to predicted atmospheric neutrino fluxes 
\cite{IceCube_PRL14}. 
A refinement of the IceCube search technique to extend the neutrino sensitivity
down to 10 TeV \cite{IceCube_PRD15}, yielded a power-law fit to the measured flux without 
cut-off given by $dN/dE = \Phi_0 (E_\nu/E_0)^{-\gamma}$
with $\Phi_0=2.06 \times 10^{-18}~{\rm GeV^{-1}~cm^{-2}~s^{-1}~sr^{-1}}$, $E_0=10^5$ GeV, and $\gamma=2.46$. 
If this flux is extrapolated to $10^{20}$ eV it would produce ${\sim} 0.1$ events in Auger.

\end{enumerate}

\section{Acknowledgments}

The successful installation, commissioning, and operation of the Pierre Auger Observatory would not have been possible without the strong commitment and effort from the technical and administrative staff in Malarg\"{u}e. 
We are very grateful to the following agencies and organizations for financial support: 

 Comisi\'{o}n Nacional de Energ\'{\i}a At\'{o}mica, 
 Fundaci\'{o}n Antorchas, Gobierno de la Provincia 
 de Mendoza, Municipalidad de Malarg\"{u}e, 
 NDM Holdings and Valle Las Le\~{n}as, in gratitude 
 for their continuing cooperation over land access, 
 Argentina; the Australian Research Council; Conselho 
 Nacional de Desenvolvimento Cient\'{\i}fico e 
 Tecnol\'{o}gico (CNPq), Financiadora de Estudos e 
 Projetos (FINEP), Funda\c{c}\~{a}o de Amparo \`{a} 
 Pesquisa do Estado de Rio de Janeiro (FAPERJ), 
 S\~{a}o Paulo Research Foundation (FAPESP) 
 Grants No. 2010/07359-6 and No. 1999/05404-3, 
 Minist\'{e}rio de Ci\^{e}ncia e Tecnologia (MCT), 
 Brazil; Grant No. MSMT-CR LG13007, No. 7AMB14AR005, 
 and the Czech Science Foundation Grant No. 14-17501S, 
 Czech Republic;  
 Centre de Calcul IN2P3/CNRS, Centre National de la 
 Recherche Scientifique (CNRS), Conseil R\'{e}gional 
 Ile-de-France, D\'{e}partement Physique Nucl\'{e}aire 
 et Corpusculaire (PNC-IN2P3/CNRS), D\'{e}partement 
 Sciences de l'Univers (SDU-INSU/CNRS), Institut 
 Lagrange de Paris (ILP) Grant No. LABEX ANR-10-LABX-63, 
 within the Investissements d'Avenir Programme  
 Grant No. ANR-11-IDEX-0004-02, France; 
 Bundesministerium f\"{u}r Bildung und Forschung (BMBF), 
 Deutsche Forschungsgemeinschaft (DFG), 
 Finanzministerium Baden-W\"{u}rttemberg, 
 Helmholtz Alliance for Astroparticle Physics (HAP), 
 Helmholtz-Gemeinschaft Deutscher Forschungszentren (HGF), 
 Ministerium f\"{u}r Wissenschaft und Forschung, Nordrhein Westfalen, 
 Ministerium f\"{u}r Wissenschaft, Forschung und Kunst, Baden-W\"{u}rttemberg, Germany; 
 Istituto Nazionale di Fisica Nucleare (INFN), Ministero dell'Istruzione, dell'Universit\'{a} 
 e della Ricerca (MIUR), Gran Sasso Center for Astroparticle Physics (CFA), CETEMPS Center 
 of Excellence, Ministero degli Affari Esteri (MAE), Italy; 
 Consejo Nacional de Ciencia y Tecnolog\'{\i}a (CONACYT), Mexico; 
 Ministerie van Onderwijs, Cultuur en Wetenschap, 
 Nederlandse Organisatie voor Wetenschappelijk Onderzoek (NWO), 
 Stichting voor Fundamenteel Onderzoek der Materie (FOM), Netherlands; 
 National Centre for Research and Development, 
 Grants No. ERA-NET-ASPERA/01/11 and 
 No. ERA-NET-ASPERA/02/11, National Science Centre,
 Grants No. 2013/08/M/ST9/00322, No. 2013/08/M/ST9/00728 
 and No. HARMONIA 5 - 2013/10/M/ST9/00062, Poland; 
 Portuguese national funds and FEDER funds within 
 Programa Operacional Factores de Competitividade 
 through Funda\c{c}\~{a}o para a Ci\^{e}ncia e a  Tecnologia (COMPETE), Portugal; 
 Romanian Authority for Scientific Research ANCS, 
 CNDI-UEFISCDI partnership projects Grants No. 20/2012 
 and No. 194/2012, Grants No. 1/ASPERA2/2012 ERA-NET, 
 No. PN-II-RU-PD-2011-3-0145-17 and No. PN-II-RU-PD-2011-3-0062, 
 the Minister of National  Education, Programme  
 Space Technology and Advanced Research (STAR), 
 Grant No. 83/2013, Romania; 
 Slovenian Research Agency, Slovenia; 
 Comunidad de Madrid, FEDER funds, Ministerio de Educaci\'{o}n y Ciencia, 
 Xunta de Galicia, European Community 7th Framework Program, 
 Grant No. FP7-PEOPLE-2012-IEF-328826, Spain; 
 Science and Technology Facilities Council, United Kingdom; 
 Department of Energy, 
 Contracts No. DE-AC02-07CH11359, No. DE-FR02-04ER41300, 
 No. DE-FG02-99ER41107 and No. DE-SC0011689, 
 National Science Foundation, Grant No. 0450696, 
 The Grainger Foundation, USA; 
 NAFOSTED, Vietnam; 
 Marie Curie-IRSES/EPLANET, European Particle Physics 
 Latin American Network, European Union 7th Framework 
 Program, Grant No. PIRSES-2009-GA-246806; and UNESCO.


\end{document}